\documentclass{ws-ijmpd}

\usepackage{graphicx}
\usepackage{euscript,amssymb}
\usepackage{amsfonts}
\usepackage{amssymb}

\begin{document}

\markboth{Mariam Bouhmadi-L\'{o}pez, Pedro F. Gonz\'{a}lez-D\'{\i}az, Prado Mart\'{\i}n-Moruno}
{On the generalised Chaplygin gas: worse than a big rip or quieter than a sudden singularity?}

\catchline{}{}{}{}{}
%
%%%%%%%%%%%%%%%%%%%%%%%%%%%%%%%%%%%%%%%%%%%%%%%%%%%%%%%%%%%%%%%%%%%%
\title{On the generalised Chaplygin gas: worse than a big rip or quieter than a sudden singularity?}

\author{Mariam Bouhmadi-L\'{o}pez}

\address{Centro Multidisciplinar de Astrof\'{\i}sica - CENTRA, Departamento de F\'{\i}sica, Instituto Superior T\'ecnico, Av. Rovisco Pais 1,
1049-001 Lisboa, Portugal
\\ mariam.bouhmadi@fisica.ist.utl.pt}

\author{Pedro F. Gonz\'{a}lez-D\'{\i}az}

\address{Colina de los Chopos, Centro de F\'{\i}sica
``Miguel A. Catal\'{a}n'', Instituto de Matem\'{a}ticas y F\'{\i}sica Fundamental,
Consejo Superior de Investigaciones Cient\'{\i}ficas, Serrano 121,
28006 Madrid, Spain\\
p.gonzalezdiaz@imaff.cfmac.csic.es}

\author{Prado Mart\'{\i}n-Moruno}

\address{Colina de los Chopos, Centro de F\'{\i}sica
``Miguel A. Catal\'{a}n'', Instituto de Matem\'{a}ticas y F\'{\i}sica Fundamental,
Consejo Superior de Investigaciones Cient\'{\i}ficas, Serrano 121,
28006 Madrid, Spain\\
pra@imaff.cfmac.csic.es}
\maketitle

\begin{history}
\received{Day Month Year}
\revised{Day Month Year}
\comby{Managing Editor}
\end{history}

\begin{abstract}
Although it has been believed that the models with generalised Chaplygin gas do not contain singularities, in a previous work we have studied how a big freeze could take place in some kinds of phantom generalised Chaplygin gas. In the present work, we study some types of generalised Chaplygin gas in order to show how different sorts of singularities could appears in such models, in the future or in the past. We point out that: (i) singularities may not be originated from the phantom nature of the fluid, and (ii) if initially the tension of the brane in a brane-world Chaplygin model is large enough then an infrared cut off appears in the past.
\end{abstract}


\section{Introduction}

In the past few years the observational data have forcefully shown that the current evolution of the universe is accelerating, \cite{Mortlock:2000zu}, so prompting a large number of theoretical models aiming at explaining these data, including quintessence scenarios, modifications of gravity, models which contain extra dimensions, etc \cite{Copeland:2006wr}. One of the most popular among these models is the generalised Chaplygin gas (GCG) which was initially introduced motivated by its connection with string theory as, furthermore it can admit a supersymmetric generalisation \cite{Kamenshchik}. In addition, GCG could explain a smooth transition from a dark-matter-dominated phase to the current epoch, as it interpolate in a natural way between dust at small scale factors and a cosmological constant on large scale factors. The possibility that the stuff which causes the accelerating expansion of the universe is a phantom fluid, i. e., a fluid with the equation of state parameter $w<-1$ (where $p=w\rho$ with $\rho>0$), has led to the consideration of the phantom GCG (PGCG) models. Although by itself a phantom fluid could produce the known big rip singularity in the future \cite{Caldwell:1999ew} it has been shown that PGCG prevents that doomsday take place \cite{Gonzalez-Diaz:2003bc,Bouhmadi-Lopez:2004me}.
Even so, PGCG models are not free from other future singularities; in fact, some of such models have been recently seen to show the so-called big freeze singularity which is characterised by a finite value of the scale factor at a given finite time, \cite{BLGDMM}. On the other hand, a dual phantom fluid can also be defined on a Randall-Sundrum type 1 scenario \cite{Yurov:2006we}. Such a fluid is characterised by an equation of state parameter $w<-1$ when the null energy condition is fulfilled. It appears therefore of interest to study the dual PGCG (DPGCG) \cite{BLGDMM} as well.
Also considered in the present paper will be the different kinds of GCG and their characteristics, emphasising those models which show past or future singularities. It is obtained that most of the studied scenarios appear plagued by distinct kinds of singularities both in the future and in the past.

This paper can be outlined as follows. In section II we analyse the different PGCG models and show their different behaviours in the future and the past, classifying the involved singularities. DPGCG models are considered in section III, checking that whereas the brane evolution describes a universe whose future evolution is similar to those considered in section II; in the past such a evolution could drastically differ, depending on the value of the brane tension. We see in section IV that even in the absence of a phantom fluid, past and future singularities could appear in the case of a GCG satisfying null, weak and strong energy conditions. In section V we summarise and conclude, adding some further comments.

\section{PGCG}\label{PGCG}

The PGCG model was originally introduced in Refs.~\cite{Bouhmadi-Lopez:2004me,Khalatnikov:2003im}. It satisfies the same equation of state as GCG \cite{Kamenshchik}, i. e.
\begin{equation}\label{uno}
p=-\frac{A}{\rho^{\alpha}},
\end{equation}
where A is a positive constant and $\alpha$ is a parameter. If $\alpha=1$, Eq.~(\ref{uno}) corresponds to the equation of state of a Chaplygin gas. The conservation of the energy momentum tensor then implies 
\begin{equation}\label{dos}
\rho=\left(A+\frac{B}{a^{3(1+\alpha)}}\right)^{\frac{1}{1+\alpha}},
\end{equation}
with $B$ a constant parameter.

In order to consider a phantom GCG (PGCG) two conditions have to be satisfied: (i) $\rho>0$ and (ii) $p+\rho<0$.
One can notice that there are four possibilities:

\begin{romanlist}[(ii)]
\item $A>0$, $B<0$ and $1+\alpha>0$.
\item $A>0$, $B<0$ and $1+\alpha<0$.
\item $A<0$,  $B>0$, $1+\alpha>0$ and  $(1+\alpha)^{-1}=2n$, with $n$ some positive integer number.
\item $A<0$,  $B>0$, $1+\alpha<0$ and  $(1+\alpha)^{-1}=2n$, with $n$ some negative integer number.
\end{romanlist}
carrying out these conditions for a fluid described by Eqs.~(\ref{uno}) and (\ref{dos}). Each of these possibilities corresponds to a different type of PGCG and can be considered as a different cosmological model. Condition (ii) for a universe whose energy density budget is dominated by  phantom energy results in  a super-inflationary expansion, in fact a super-accelerating  phase. The different sorts of PGCG fulfil the conservation of the energy momentum tensor.

\subsection{PGCG I}\label{I}

For this model $A>0$, $B<0$ and $1+\alpha>0$. This fluid has been studied in great detail in Ref.~\cite{Bouhmadi-Lopez:2004me} and we review here the characteristics of a universe filled with it, both in the past and in the future.

From Eq.~(\ref{dos}), we can realise that the cosmic scale factor takes values in the interval $a_{{\rm min}}\leq a<\infty$ which corresponds to $0\leq\rho<A^{1/(1+\alpha)}$, where
\begin{equation}\label{tres}
a_{{\rm min}}=\left|\frac{B}{A}\right|^{\frac{1}{3(1+\alpha)}}.
\end{equation}
Then, in this model the scale factor grows up from a non-vanishing value to infinity, whereas the energy density increases as the universe expands until it reaches  its asymptotic value $A^{1/(1+\alpha)}$. Throughout this paper, we use units such that $\frac{\kappa_4^2}{3}=1$ and $\hbar=c=1$, where $\kappa_4^2$ is the gravitational constant. As it is shown in Ref.~\cite{Bouhmadi-Lopez:2004me}, we can integrate analytically  the Friedmann equation (where here and in what follows, we choose the flat chart to describe the geometry of a homogeneous and isotropic universe) to note that the cosmic time required by the scale factor to be infinitely large is infinite. Therefore, 
the universe would avoid a big rip singularity \cite{Caldwell:1999ew}, even if it is filled with phantom energy \cite{Gonzalez-Diaz:2003bc,Bouhmadi-Lopez:2004me}. Hence, the universe would  expand infinitely in the future cosmic time, reaching an infinite size with a finite asymptotic energy density; i.e the universe would be asymptotically de Sitter \cite{Bouhmadi-Lopez:2004me}.

The time elapsed since the origin of the universe to a given scale factor close to its initial value can be easily estimated by analysing the behaviour of the model in the neighbourhood of $a_{{\rm min}}$. For $a_0$ (at a given cosmic time $t_0$) close to but still larger than $a_{{\rm min}}$, the scale factor dependence on the cosmic time, $t$, can be approximated by
\begin{equation}\label{cuatro}
a\simeq a_{{\rm min}}\left\{1+\left[\left(\frac{a_0}{a_{{\rm min}}}-1\right)^{\frac{1+2\alpha}{2(1+\alpha)}}+\frac{1+2\alpha}{2(1+\alpha)}\left(3A(1+\alpha)\right)^{\frac{1}{2(1+\alpha)}}(t-t_0)\right]^{\frac{2(1+\alpha)}{1+2\alpha}}\right\},
\end{equation}
for $\alpha\neq -1/2$ and by
\begin{equation}
a\simeq a_{{\rm min}}\left\{1+\left(\frac{a_0}{a_{\rm min}}-1\right)\exp\left[\frac{3}{2}A\left(t-t_0\right)\right]\right\},
\label{cuatro-2}\end{equation}
for $\alpha=-1/2$.
Then, if $-1<\alpha\leq -1/2$, the universe ought to have been born an infinite time ago, avoiding the standard big bang singularity\footnote{Although there
is no standard big bang singularity as the scalar curvature is well
defined, the model is singular in curvature for a non cosmological
observer \cite{FernandezJambrina:2007sx}.  We thank  L. Fern\'andez-Jambrina
 for pointing out this to us.}. In fact, both the energy density and the pressure vanish at the infinite past. On the other hand, if $-1/2<\alpha$ we can also estimate the time interval since the universe was born until it reaches a size with scale factor $a_0$, not much larger than $a_{{\rm min}}$, to be
\begin{equation}\label{cinco}
t_0-t_{{\rm min}}\simeq\left(\frac{1}{3A(1+\alpha)}\right)^{\frac{1}{2(1+\alpha)}}\frac{2(1+\alpha)}{1+2\alpha}\left(\frac{a_0}{a_{{\rm min}}}-1\right)^{\frac{1+2\alpha}{2(1+\alpha)}}.
\end{equation}
The energy density, but not necessarily the pressure, is again well defined at $t_{{\rm min}}$ (where it vanishes). The question becomes now appropriate, does the universe face a singularity at this minimum scale factor? This is the question we tackle next. Using the Friedmann equation, the conservation law of a perfect fluid and Eq.~(\ref{dos}), the time derivative of the Hubble parameter, $\dot H$, and the scalar curvature for a GCG can be written as
\begin{equation}\label{H}
\dot{H}=-\frac{3}{2}(p+\rho),
\end{equation}
\begin{equation}\label{R}
R=12\left(A+\frac{B}{4a^{3(1+\alpha)}}\right)\left(A+\frac{B}{a^{3(1+\alpha)}}\right)^{-\frac{\alpha}{1+\alpha}}.
\end{equation}
Consequently, if $0<\alpha$ the pressure, and both these two quantities blow up, i.e.  there is  a past sudden singularity \cite{Barrow:2004xh}, a big brake singularity \cite{Gorini} or a Type II singularity in the notation of Ref.~\cite{Nojiri}, with the peculiarity that the energy density vanishes rather than acquiring a positive non-vanishing value. But if $-1/2<\alpha<0$, even though $R$, $H$, $\dot{H}$ and the pressure are all well defined at $a_{{\rm min}}$, higher derivatives of the Hubble parameter could still diverge. In fact, the $n$-th derivative of the Hubble parameter respect to the cosmic time has the following structure
\begin{equation}\label{H2}
\frac{d^n}{dt^n}H\propto\rho^{-n(\frac12 +\alpha)+\frac12}+C_1\rho^{-(n-1)(\frac12 +\alpha)+1}
+\textrm{higher powers in  } \rho,
\end{equation}
where $C_1$ is a constant. From this expression, we conclude that the $n$-th derivative diverges at $a_{{\rm min}}$ if $1/(1+2\alpha)<n$, unless $\alpha$ can be written as $\alpha=1/(2p)-1/2$, where $p$ is a positive integer. For a given $\alpha(\neq 1/(2p)-1/2)$, the n-th derivative of $H$ where $n=1+E(1/(2\alpha +1))$ reaches very large values at $a_{{\rm min}}$.  In the previous expression $E$ denotes the integer value function. This corresponds to a type IV singularity in the terminology of Ref.~\cite{Nojiri}. In addition, this singularity is quieter than a sudden singularity as the energy density and pressure are well defined at the singularity, as $H$ and its first time derivative also are, but higher derivatives of the Hubble parameter can blow up.
In the particular case $\alpha=1/(2p)-1/2$, where $p$ is an integer, all the derivatives of the Hubble parameter respect to the cosmic time are always well defined, however. The reason is the following: the $1,..,(p-1)$-th derivatives are well behaved at $t_{\rm{min}}$ (see Eq.~(\ref{H2})), the next derivative, which reads
\begin{eqnarray}
\frac{d^p}{dt^p}H\propto C_2+ \rho^{\frac{p+1}{2p}}  +\textrm{higher powers in  } \rho,
\end{eqnarray}
with $C_2$ a constant, which is also well defined at the minimum size of the universe.
Then
\begin{eqnarray}
\frac{d^{p+1}}{dt^{p+1}}H\propto \rho^{1/2}  +\textrm{higher powers in  } \rho.
\end{eqnarray}
Therefore, the lower order of $\rho$ in the (p+1)-th derivative is similar to the order of $\rho$ in $H$ and so well defined, too. Thus, we are back to our starting point. We can then conclude that all  the derivatives of $H$ are well defined at $t_{\rm{min}}$ if $\alpha=1/(2p)-1/2$ being $p$ a positive integer. 

In summary, if $-1<\alpha\leq-1/2$, then there is no singularity at a finite time in the past. If $-1/2<\alpha$ and $\alpha\neq 1/(2p)-1/2$, where $p$ is an integer, there is a singularity at $a_{{\rm min}}$ which  can be of two different types depending on (i) $0<\alpha$ or (ii) $-1/2<\alpha<0$. In the former case, i.e. $0<\alpha$, it corresponds to a sudden past singularity \cite{Barrow:2004xh} with a vanishing energy density, while in the latter case, $-1/2<\alpha<0$, it corresponds to a divergence in higher derivatives of the Hubble parameter (in fact, a type IV singularity\footnote{\label{footnote1}
It is a singularity as far as the derivatives of the Hubble parameter diverge, even though the scalar curvature and Riemann tensor do not.} in the notation of Ref.~\cite{Nojiri}).

\subsection{PGCG II}\label{II}

\begin{figure}[t]
\begin{center}
\includegraphics[width=6cm]{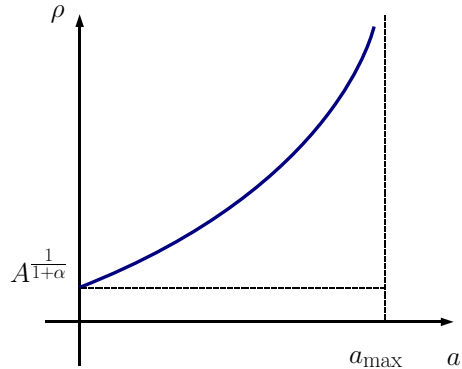}
\end{center}
\caption{A schematic plot of a future big freeze singularity. This singularity takes places at a finite future cosmic time where not only the energy density but also the pressure blows up. Therefore, both the Hubble rate and its cosmic time diverge as well at the big freeze event.}
\label{fbigfree}
\end{figure}

Now, we consider a PGCG like in the previous case; i.e.  $A>0$ and $B<0$, but where $1+\alpha<0$ \cite{Bouhmadi-Lopez:2004me,Sen:2005sk,BLGDMM}. Here the scale factor is bounded $0\leq a\leq a_{{\rm max}}$ and the energy density is larger than $A^{1/(1+\alpha)}$, where
\begin{equation}\label{seis}
a_{\rm{max}}=\left|\frac{B}{A}\right|^{\frac{1}{3(1+\alpha)}}.
\end{equation} 
The cosmic time scales with the scale factor as
\begin{eqnarray}
t_{\rm{max}}-t=&-&\frac{2}{3(1+2\alpha)}A^{-\frac{1}{2(1+\alpha)}}\left[1-\left(\frac{a}{a_{\rm{max}}}\right)^{-3(1+\alpha)}\right]^{\frac{1+2\alpha}{2(1+\alpha)}}\nonumber \\  &\times& {\rm{F}}\left(1,\frac{1+2\alpha}{2(1+\alpha)};\frac{3+4\alpha}{2(1+\alpha)};
1-\left(\frac{a}{a_{\rm{max}}}\right)^{-3(1+\alpha)}\right),
\end{eqnarray}
where ${\rm{F}}(b,c;d;e)$ is a hypergeometric function \cite{Gradshteyn}. In the previous expression $t_{\rm{max}}$ corresponds to the cosmic time when a universe filled by this fluid would  hit a big freeze singularity \cite{BLGDMM}; i.e. a singularity at a finite scale factor, $a_{\rm{max}}$, and cosmic time, $t_{\rm{max}}$, and  where both the energy density and pressure blow up, as the Hubble parameter and its cosmic derivative  \cite{BLGDMM} (see Fig.~\ref{fbigfree}) do as well.
It can be easily checked  that the cosmic time since the universe has a given size (at a given $t$) till it hits the big freeze is finite\footnote{\label{series}
A hypergeometric series $\textrm{F}(b,c;d;e)$, also called a hypergeometric function, converges at any value $e$ such that $|e|\leq 1$, whenever $b+c-d<0$. However, if  $0 \leq  b+c-d < 1$ the series does not converge at $e=1$. In addition, if  $1 \leq b+c-d$, the hypergeometric function blows up at $|e|=1$ \cite{Gradshteyn}.}
. This result confirms  our previous approximation close to $a_{\rm{max}}$ \cite{BLGDMM}
\begin{equation}%\label{siete}
a\simeq a_{\rm{max}}\left\{1-\left[\frac{1+2\alpha}{2(1+\alpha)}\right]^{\frac{2(1+\alpha)}{1+2\alpha}}A^{\frac{1}{1+2\alpha}}
|3(1+\alpha)|^{\frac{1}{1+2\alpha}}(t_{\rm{max}}-t)^{\frac{2(1+\alpha)}{1+2\alpha}}\right\}.
\label{amaxtmax}\end{equation}
It can also be noticed that when the scale factor reaches its maximum finite size, the scalar curvature blows up (see Eq.~(\ref{R})). This singularity can be classified as Type III in the notation of Ref.~\cite{Nojiri} (see also Refs.~\cite{NOT2,NOT3}). In this model the universe expands until it reaches a maximum size, at which moment it freezes with an infinite energy density; for this reason this doomsday has been named big freeze singularity \cite{BLGDMM}.

%%%%%%%%%%%%%%%FIG1%%%%%%%%%%%%%%%%%%%%%%%%%%%%%%%%%%%%%

%%%%%%%%%%%%%%%%%%%%%%%%%%%%%%%%%%%%%%%%%%%%%%%%%%%%%%%%

Finally, let us briefly point out  that  a universe filled with this kind of PGCG  is asymptotically de Sitter in the past. The scale factor in the regime of small scale factor can be approximated by 
\begin{equation}\label{ocho}
a\simeq a_0\exp\left(A^{\frac{1}{2(1+\alpha)}}(t-t_0)\right),
\end{equation}
where $a_0$ is a non-vanishing small scale factor. Therefore, the universe starts its evolution at an infinite past ($a\rightarrow0$ as $t-t_0\rightarrow\infty$) where there is no big bang singularity.

\subsection{PGCG III and IV}\label{III}

These PGCG  correspond to having  $A<0$ and $B>0$ in Eqs.~(\ref{uno})-(\ref{dos}). Hence, in order to satisfy the phantom energy conditions; i.e. $\rho>0$ and $p+\rho<0$, the parameter $\alpha$ must be quantised so that $(1+\alpha)^{-1}=2n$, with $n$ an integer. There are two kinds of phantom fluids depending on $n$; i.e. $(1+\alpha)^{-1}$,  positive or negative. We will refer to the former fluid ($n>0$) as the phantom GCG III and to the latter ($n<0$) as the  phantom GCG IV.

A universe filled with a type III phantom GCG would have a minimum scale factor ($a_{{\rm min}}\leq a$, where $a_{{\rm min}}$ is given by Eq.~(\ref{tres})) at an infinite past which  can be noticed by using the conveniently modified Eqs.~(\ref{cuatro}) and (\ref{cuatro-2}) when $A$ is replaced by $|A|$. As it can be realised, the advantage of having a  quantised  $\alpha$ is to remove the possible different past singularities (at a finite cosmic time) present in a phantom GCG~I model. After all, this advantage is not that surprising as having  $n>0$ implies $-1<\alpha\leq -1/2$, precisely a situation without past singularities (at a finite cosmic time) for the phantom GCG I . Finally, we note that such a universe would be asymptotically de Sitter in the future.

On the other hand, a universe filled with a type IV phantom GCG would have a maximum scale factor ($a\leq a_{{\rm max}}$, where $a_{{\rm max}}$ is given by Eq.~(\ref{seis})). Then, the universe would hit a big freeze singularity at a finite future cosmic time (Eq.~(\ref{amaxtmax}) after replacing $A$ for $|A|$ applies in this case), where the energy density and  pressure would diverge (see Fig.~\ref{fbigfree}). Therefore, unlike the previous case, the quantisation of the parameter $\alpha$ is useless to remove the big freeze singularity. Let us point out that such a universe would be asymptotically de Sitter in the past and born  an infinite time ago.

A summary of the results found on this section are shown in table~I.

{\vspace*{-0.1cm}{
\begin{table}[t]
\tbl{This table summarises the asymptotic behaviour of a universe filled with each type of a phantom GCG. 
The past asymptotic behaviours labelled by (1) and (3) correspond to  $-1<\alpha\leq-1/2$ and $0<\alpha$, respectively. 
The  past asymptotic behaviour labelled by (2) corresponds to $-1/2<\alpha< 0$, where $\alpha$ cannot be expressed as $\alpha=1/(2p)-1/2$, with $p$ a positive integer. If  $-1/2<\alpha< 0$ and $\alpha$ can be expressed as $\alpha=1/(2p)-1/2$, with $p$ a positive integer, there is no past singularity and the universe is born at a finite past.}
{\begin{tabular}{|c|c|c|c|c|c|}%\toprule
\cline{1-6}
A,B	&  $1+\alpha$	&$a$	&$\rho$	                          &Past	                               &Future\\
\cline{1-6}
	&	        &	&                   &(1) infinite  past  &      \\
\cline{5-5}
$A>0$ & positive	&$a_{{\rm min}}\leq a <\infty$	&  $0\leq\rho\leq A^{1/(1+\alpha)}$                 &(2)  type IV singularity           &asymptotically de Sitter \\
\cline{5-5}
&	&	&	&(3) sudden singularity	& \\
\cline{2-6}
$B<0$	& negative	&$0\leq a<a_{{\rm max}}$	&$A^{1/(1+\alpha)}\leq\rho<\infty$	&asymptotically de Sitter/infinite  past	&big freeze singularity\\
\cline{1-6}
$A<0$	&$(2n)^{-1}>0$	&$a_{{\rm min}}\leq a <\infty$	&$0\leq\rho\leq |A|^{1/(1+\alpha)}$	&infinite  past	&asymptotically de Sitter\\
\cline{2-6}
$B>0$	&$(2n)^{-1}<0$	&$0\leq a<a_{{\rm max}}$	&$|A|^{1/(1+\alpha)}\leq\rho<\infty$	&asymptotically de Sitter/infinite  past 	&big freeze singularity\\
\cline{1-6}
\end{tabular}\label{table1}}
\end{table}}}

\section{DPGCG}\label{DPGCG}

In this section, we  thoroughly analyse the cosmology of a dual phantom GCG (DPGCG) \cite{BLGDMM}. Such a dual fluid is defined as  a perfect fluid  which satisfies  the  equation of state (\ref{uno}) with  $w<-1$ and fulfils the null energy condition \cite{Yurov:2006we,BLGDMM,Visser:1995cc}.  This definition of dual has not to be confused with the one used in Ref.~\cite{NOT3}. 

The  key point about the dual of a phantom energy  setup is that it may supply an alternative for dark energy models in theories with modified Friedmann equation like the brane-world scenario \cite{review}, despite its energy density is negative\footnote{\label{footnote2} A negative dark energy component has been also used in Ref.~\cite{Grande:2006nn} but in different set up.}. We next review how this can be the case \cite{Yurov:2006we,BLGDMM} in a Randall-Sundrum model  with a unique brane (RS1) \cite{RS}. The modified Friedmann equation on the brane reads \cite{friedmannbrane}
\begin{equation}
H^2=\rho\left(1+\frac{\rho}{2\lambda}\right),
\label{friedmannrs}
\end{equation}
where $\lambda$ is the positive brane tension. Therefore, the square of the Hubble parameter in RS1 scenario is well defined even for a negative energy density $\rho$ as long as $\rho<-2\lambda$; i.e. as long as the effective energy density of the brane is positive. In this setup, the brane is super-inflating; i.e. $\dot H >0$ where
\begin{equation}
\dot H=-\frac32 (p+\rho)\left(1+\frac{\rho}{\lambda}\right)>0.
\label{dotHrs}
\end{equation}
Consequently, a brane filled with the dual of a phantom energy is always accelerating; i.e. $\ddot a >0$, in particular if it is filled with DPGCG. Moreover, the brane is super-inflating although the dual of the phantom energy satisfies
the null energy condition; i. e. $p+\rho>0$, because $\rho$ is negative and smaller than $-2\lambda$. We remind that here the effective energy density of the brane is positive and is well behaved.

We will analyse three different types of DPGCG characterised by\footnote{The families of duals of the phantom GCG is much larger and can be characterised by: 
\begin{eqnarray}
1+\alpha &=& \frac{1+2(m+n)}{1+2n}, \,\,\, m,n\in \mathbb{Z}, \,\,\, A<0, B>0,\\
\rho&=&-\left(|A|-\frac{B}{a^{\frac{3\left[1+2(m+n)\right]}{1+2n}}}\right)^{\frac{1+2n}{1+2(m+n)}}<0,\,\,\, \textrm{with}\,\,\, -|A|+\frac{B}{a^{\frac{3[1+2(m+n)]}{1+2n}}}<0, \nonumber \\%{\label{densitygeneral}} 
\\
p&=&|A|\left(|A|-\frac{B}{a^{\frac{3[1+2(m+n)]}{1+2n}}}\right)^{-\frac{2m}{1+2(m+n)}}>0.
%\label{pressuregeneral}
\end{eqnarray}
Here we restrict to the case $m=-n$. These set of solutions fulfil the conservation of the energy momentum tensor.} 
\begin{romanlist}[(b)]

\item $A<0$,  $B>0$, $1+\alpha>0$ and  $(1+\alpha)^{-1}=2n +1$, with $n$ a positive integer,

\item $A<0$,  $B>0$, $1+\alpha<0$,  $(1+\alpha)^{-1}=2n +1$, with $n$ a negative integer, and $|A|^{1/(1+\alpha)}<2\lambda$,

\item $A<0$,  $B>0$, $1+\alpha<0$,  $(1+\alpha)^{-1}=2n +1$, with $n$ a negative integer, and $2\lambda\leq |A|^{1/(1+\alpha)}$,
\end{romanlist}
which we will refer to as DPGCG I, II and III, respectively. As it can be noticed a commonness of DPGCGs is that the parameter $\alpha$ is quantised; more precisely, $1+\alpha=1/(1+2n)$, where $n$ is an integer. 

Our main objective in what follows of this section is to analyse the cosmological evolution of the brane and, in particular, the singularities that may arise in a brane filled with DPGCG in the RS1 model\footnote{An anisotropic brane filled with a Chaplygin gas has been analysed in Ref.\cite{Harko}.}.

\subsection{DPGCG I}

\begin{figure}[h]
\begin{center}
\includegraphics[width=8cm]{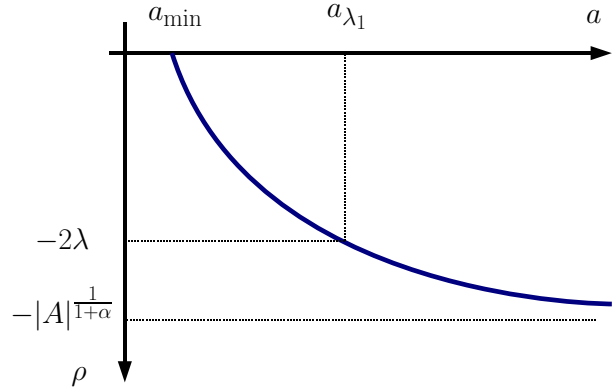}
\end{center}
\caption{This plot shows the energy density $\rho$ of DPGCG~I versus the scale
factor of the brane. The brane tension $\lambda$ acts as a cut off on the
energy density at early time, therefore allowing only values of the scale factor larger
than $a_{\lambda_1}$, where $\rho$ equates $-2\lambda$. At large values of the scale factor
the DPGCG~I energy density approaches a constant negative energy density
for which DPGCG~I mimics a negative cosmological constant, while the
brane is asymptotically de Sitter.}
\label{dpgcg1}
\end{figure}

We consider a brane filled with DPGCG I. In this model, the minimum energy density  must be smaller than $-2\lambda$ (see Fig.~\ref{dpgcg1}); i.e. $2\lambda< |A|^{1+2n}$, in order for the brane to have a  Lorentzian evolution. This feature implies that the expansion of the brane starts  at a non-vanishing scale factor $a_{\lambda_1}$ defined\footnote{\label{footnote3} The analysis of this section is restricted to the expanding phase of the  brane.} as
\begin{equation}\label{diez}
a_{\lambda_1}=a_{{\rm min}}\left[1-(2\lambda)^{\frac{1}{1+2n}}|A|^{-1}\right]^{-\frac{1+2n}{3}},
\end{equation}
where the Hubble rate vanishes and $\rho=-2\lambda$. In the neighbourhood of $a_{\lambda_1}$, the scale factor growth can be approximated by 
\begin{equation}\label{once}
a\simeq a_{\lambda_1}\left\{1+\frac34 (2\lambda)^{\frac{2n}{1+2n}}
|A|\left[1-(2\lambda)^{\frac{1}{1+2n}}|A|^{-1}\right](t-t_{\lambda_1})^2\right\},
\end{equation}
which implies that the brane starts evolving smoothly, without any big bang singularity. Moreover,  the brane is born at a finite past cosmic time $t_{\lambda_1}$ and a given radius $a_{\lambda_1}$, which is set up by the brane tension for a given DPGCG I, that should not be confused with $a_{\rm{min}}$ given by Eq.~(\ref{tres}),  due to the fact that the brane tension acts as an early time energy cut off on the DPGCG I energy density. In other words, the brane tension excludes those radii of the brane such that $a_{\rm{min}}<a<a_{\lambda_1}$, where the energy density vanishes at $a_{\rm{min}}$. This cut off effect implies that the brane starts evolving at a finite past cosmic time as measured by an observer confined on the brane.

Later on, the brane keeps expanding super-inflationarily (cf. Eq.~(\ref{dotHrs})) until the Hubble parameter approaches its maximum value $H_i$ (or $\rho$ acquires its minimum value; notice that $2\lambda\leq|\rho|<|A|^{{1+2n}}$), where
\begin{equation}
H_{i}=\sqrt{|A|^{{1+2n}}\left(-1+\frac{|A|^{{1+2n}}}{2\lambda}\right)},
\label{Hi}\end{equation}
at large scale factor. It turns out that the brane is asymptotically  de Sitter in the future despite that the DPGCG~I energy density is negative and mimics a negative cosmological constant at large scale factor. At this respect, we recall that the modified Friedmann equation on the brane  is quadratic on the total energy density of the brane (cf. Eq.~(\ref{friedmannrs})) and therefore the effective energy density on the brane is positive . 

%%%%%%%%%%%%%%%%FIG2%%%%%%%%%%%%%%%%%%%%%%%%%%

%%%%%%%%%%%%%%%%%%%%%%%%%%%%%%%%%%%%%%%%%%%%%%%%%%%%%%%%%%%%%%%%%%%%%%%%%%%%

\subsection{DPGCG~II}

We move on to analyse what happens to a brane filled with a DPGCG~II. First of all, let us point out that the main difference respect to the previous case is that the brane faces a big freeze singularity at a finite future cosmic time. Therefore, the  fulfilment of the null energy condition \cite{Visser:1995cc} does  not avoid the happening of such a singularity.

\begin{figure}[h]
\begin{center}
\includegraphics[width=8cm]{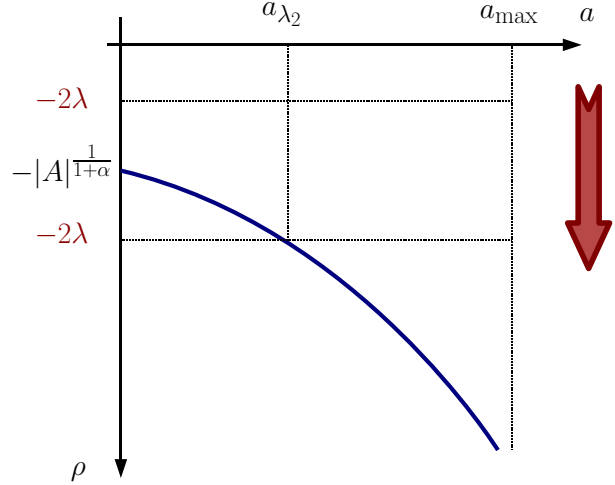}
\end{center}
\caption{This plot shows the energy density $\rho$ of DPGCG~II and DPGCG~III versus the scale
factor of the brane. The late time behaviour of a brane filled with any of these fluids is the same and shows a big freeze singularity. However, the early time evolution is different for the two fluids. For DPGCG~III the brane tension $\lambda$ does not act as a cut off on the energy density (at early time) as $2\lambda\leq|A|^{\frac{1}{1+\alpha}}$. The reason is that the brane tension is too small. As the brane tension is increased (represented schematically by an arrow on the plot), the system behaves as a DPGCG~II, where now $\lambda$ acts as a cut off on the energy density. In this case, the brane is allowed to have only values for the scale factor on the range  $a_{\lambda_2}\leq a <a$.}
\label{dpgcg2}
\end{figure}

In this model, the scale factor is bounded  $a_{\lambda_2}\leq a<a_{\rm{max}}$, where  $a_{\rm{max}}$ is  defined in Eq.~(\ref{seis}) and $a_{\lambda_2}$ reads
%%%%
\begin{equation}
a_{\lambda_2}=a_{\rm{max}}\left[1-(2\lambda)^{\frac{1}{1+2n}}|A|^{-1}\right]^{-\frac{1+2n}{3}}.\label{alambda2}
\end{equation}
%%%%
The brane starts its evolution at a non-vanishing value of the scale factor $a_{\lambda_2}$ because the initial energy density of DPGCG~II is larger than the energy scale fixed by  the brane tension; i.e., $|A|^{1/(1+\alpha)}<2\lambda$ (see Fig.~\ref{dpgcg2}). Therefore, $\lambda$ acts as an early time energy cut off on $\rho$ and excludes the radii $0<a<a_{\lambda_2}$ where $-2\lambda<\rho$. Near $a_{\lambda_2}$,  the dependence of the scale factor on the brane proper time reads
%%%%%
\begin{equation}
a(t)\simeq a_{\lambda_2}\left\{1+\frac{3}{4}(2\lambda)^{\frac{2n}{1+2n}}|A|\left[1-|A|^{-1}(2\lambda)^{\frac{1}{1+2n}}\right](t-t_{\lambda_2})^2\right\},
\end{equation}
%%%%%
$t_{\lambda_2}$ being the cosmic time at which the brane has its minimum radius. Unlike a brane filled with DPGCG~III, where $2\lambda\leq |A|^{1/(1+\alpha)}$, here the brane tension prevents the existence of an infinite past for the brane. On the other hand, the Hubble rate grows so fast that the brane faces a big freeze singularity at $a_{\rm{max}}$. In fact, close to this doomsday, the cosmological expansion of the brane can be expressed as
%%%%
\begin{equation}
a(t)\simeq a_{\rm{max}}\left[1-\left(\frac{2}{\lambda}\right)^{-\frac{1}{4n}}|A|^{-\frac{1+2n}{2n}}|n|^{-\frac{1}{2n}}\left(\frac{|1+2n|}{3}\right)^{\frac{1+2n}{2n}}(t_{\rm{max}}-t)^{-\frac{1}{2n}}\right],
\end{equation}
%%%%
where the big freeze takes place at $t_{\rm{max}}$ and therefore both the energy density and pressure of DPGCG~III blow up as well as the Hubble rate and its cosmic derivative do. The happening of this singularity can be understood as a consequence of the super-inflationary expansion of the brane (cf. Eq.~(\ref{dotHrs})) and the unboundness of $|\rho|$ ($2\lambda\leq|\rho|$). Moreover, the Hubble parameter grows at high energy; i.e. close to $a_{\rm{max}}$, and for a given constant $\alpha$ is enhanced by the brane effect respect to the standard 4D situation represented by PGCG~II and the PGCG~IV fluids. Therefore, the big freeze singularity would take place in a more virulent way on the brane as $H$ would diverges at a greater rate.

%%%%%%%%%FIG3%%%%%%%%%%%%%%%%%%%%
%%%%%%%%%%%%%%%%%%%%%%%%%%%%%%%%%

\subsection{DPGCG~III}

The late time cosmology of a brane filled with DPGCG~II or DPGCG~III is the same and hence we refer the reader to see the previous subsection for a more detailed analysis. However, the early time evolution of the brane is quite different (see Fig.~\ref{dpgcg2}). In fact, the brane tension does not exclude those radii banned by DPGCG~II and the scale factor can
take any  value on $0\leq a <a_{\rm{max}}$. The brane is then allowed to have very small radii because $2\lambda\leq |A|^{1/(1+\alpha)}$  and therefore $\rho<-2\lambda$, a fact  implying a well defined Hubble parameter  even at very small scale factors (cf. Eq.~(\ref{friedmannrs})). In other words, the energy scale set up by the brane tension is so small that it does not imply any cut off on $\rho$. It turns out that  the brane evolution starts at an infinite time in the past. In fact, at small scale factor, the Friedmann equation implies
%%%
\begin{equation}
a\simeq a_0{\rm exp}\left(H_i(t-t_0)\right),
\end{equation}
%%%
if $2\lambda<|A|^{1/(1+\alpha)}$. In the previous expression $H_i$ corresponds to the primordial Hubble parameter and is defined by Eq.~(\ref{Hi}) and $a_0$ stands up for a small scale factor. Therefore, the brane is asymptotically de Sitter in the past. On the other hand, for the limiting case $2\lambda=|A|^{1/(1+\alpha)}$, the early time evolution is different (a sort of power law inflation)
%%%
\begin{equation}
a\simeq a_{\rm{max}}\left[
\left(\frac{a_0}{ a_{\rm{max}}}\right)^{\frac{3}{2(1+2n)}}
-\frac32\sqrt{\frac{2\lambda}{|1+2n|}}(t-t_0)\right]^{\frac{2(1+2n)}{3}},
\end{equation}
%%%
however, the brane has still an infinite past. We remind at this respect that $1+2n$ is negative. A summary of the results of the present section is given in table II.

\begin{table}[h]
\tbl{This table summarises the asymptotic behaviour of a universe filled with each of the different types of DPGCG which were analysed in the text.}
{\begin{tabular}{|c|c|c|c|c|c|c|}
\cline{1-7}
A,B	&$1+\alpha$	& $\lambda$&$a$	&$\rho$	&Past	&Future\\
\cline{1-7}
$A<0$	&$(1+2n)^{-1}>0$ & $2\lambda< |A|^{1/(1+\alpha)}$	&$a_{\lambda_1}\leq a<\infty$	&$2\lambda\leq|\rho|\leq |A|^{1/(1+\alpha)}$	&finite	past& asymptotically de Sitter\\
\cline{2-7}
	& & $2\lambda< |A|^{1/(1+\alpha)}$	&$0\leq a<a_{{\rm max}}$	&$|A|^{1/(1+\alpha)}\leq|\rho|<\infty$	&asymptotically de Sitter	&big freeze singularity\\
\cline{3-7}
$B>0$	& $(1+2n)^{-1}<0$ & $2\lambda= |A|^{1/(1+\alpha)}$	&$0\leq a<a_{{\rm max}}$	&$|A|^{1/(1+\alpha)}\leq|\rho|<\infty$	&power law/infinite past	&big freeze singularity\\
\cline{3-7}
&	&$|A|^{1/(1+\alpha)}<2\lambda$ &$a_{\lambda_2}<a<a_{{\rm max}}$	&$2\lambda\leq|\rho|<\infty$	&finite	past &big freeze singularity\\
\cline{1-7}
\end{tabular}\label{table2}}
\end{table}

\section{The plain GCG with singularities}
\label{GCG}

One could think that the singularities appearing in section \ref{PGCG}, in particular the big freeze one, could be a consequence from the use of a phantom fluid, since such a stuff violates the null, strong and  weak energy conditions\footnote{\label{footnote4} A similar statement would apply to a dual PGCG as it does not satisfy the weak energy condition. Notice that a dual PGCG amazingly fulfils the strong energy condition.}  \cite{Visser:1995cc}. We next show that even for GCG fulfilling such energy conditions a big freeze, a sudden or other type of singularities  may appear. This can be the case for the following set of GCG parameters:

\begin{romanlist}[(ii)]
\item $A<0$, $B>0$ and $1+\alpha>0$.
\item $A<0$, $B>0$ and $1+\alpha<0$.
\item $A>0$,  $B<0$, $1+\alpha>0$ and  $(1+\alpha)^{-1}=2n$, with $n$ some positive integer number.
\item $A>0$,  $B<0$, $1+\alpha<0$ and  $(1+\alpha)^{-1}=2n$, with $n$ some negative integer number.

\end{romanlist}
On the other hand, we will also show that the dominant energy condition \cite{Visser:1995cc} is however not fulfilled at the big freeze singularity. We present a  summary of the results found on this section in table~III.

\subsection{GCG I}\label{A}

We start analysing GCG I which is characterised by  $A<0$, $B>0$ and $1+\alpha>0$ in Eqs.~(\ref{uno}) and (\ref{dos}).
This  fluid behaves like a ``dust'' for small scale factor \footnote{\label{footnote5}
Although, $p/\rho \rightarrow 0$ the pressure can diverge at small scale factors if $-1<\alpha<0$. On the other hand, if $\alpha$ is positive, the pressure vanishes when the scale factor is small.}; i.e. $p/\rho \rightarrow 0$. Moreover, the universe starts evolving with an infinite Hubble parameter and an infinite deceleration parameter ($\ddot a$ is very large and negative). Afterwards, the universe starts expanding, although never accelerating, until it reaches its maximum size given by Eq.~(\ref{seis}).
Now, how long does it take for such a universe to reach its maximum size? It can be easily checked that close to $a_{max}$ the scale factor growth can be approximated by
\begin{equation}
a\simeq a_{{\rm max}}\left\{1-\left[\left(1-\frac{a_0}{a_{\rm{max}}}\right)^{\frac{1+2\alpha}{2(1+\alpha)}}-{\frac{1+2\alpha}{2(1+\alpha)}}
\bigg(3|A|(1+\alpha)\bigg)^{\frac{1}{2(1+\alpha)}}(t-t_0)
\right]^{\frac{2(1+\alpha)}{1+2\alpha}} 
\right\},
\label{amaxappr1}
\end{equation}
for $\alpha\neq -1/2$ and by
\begin{equation}
  a\simeq a_{{\rm max}}\left\{1-\left(1-\frac{a_0}{a_{\rm{max}}}\right)\exp\left[-\frac32 |A|(t-t_0)\right]\right\}
\label{amaxappr2}
\end{equation}
if $\alpha= -1/2$. In the previous equations $a_0=a(t_0)$ is a scale factor close to $a_{{\rm max}}$.

Consequently,  if $-1<\alpha\leq -1/2$, the universe would reach its maximum size in an infinite cosmic time. Then the universe does not hit a future singularity. In fact, the scalar curvature is well defined and vanishes when the universe reaches its maximum size where the energy density and the pressure of GCG both vanish.

However, for $-1/2<\alpha$ the universe would reach its maximum size in a finite cosmic time in the future
\begin{equation}
t_{{\rm max}}-t_0=\frac{2(1+\alpha)}{1+2\alpha}\left[3|A|(1+\alpha)\right]^{-\frac{1}{2(1+\alpha)}}\left(1-\frac{a_0}{a_{{\rm max}}}\right)^{\frac{1+2\alpha}{2(1+\alpha)}},
\end{equation}
where it will face a curvature singularity if $\alpha$ is positive. Indeed, the scalar curvature blows up (see Eq.~(\ref{R})) as well as it does the pressure of GCG. Therefore, there is  a sudden singularity  \cite{Barrow:2004xh},  a big brake singularity \cite{Gorini} or a Type II singularity in the notation of Ref.~\cite{Nojiri}, at $t=t_{\rm{max}}$, with the peculiarity that the energy density vanishes at the singularity rather than approaching a positive value. If $-1/2<\alpha<0$, the situation looks a bit more complicated. In this case, although the GCG energy density and the pressure are well defined at $t_{{\rm max}}$, and consequently also $H$ and $\dot H$ do, it turns out that higher derivatives of the Hubble parameter can diverge when the universe reaches its maximum size. The $n$-th derivative of the Hubble parameter respect to the cosmic time has the structure given  in Eq.~(\ref{H2}). Therefore, following a reasoning similar to the one used in subsection \ref{I}, it can  be straightforwardly shown that there always are higher derivatives of $H$ that diverges at $t=t_{max}$, unless $\alpha$ can be written as $\alpha=1/(2p)-1/2$, where $p$ is a positive integer. This corresponds to a Type IV singularity in the notation of Ref.~\cite{Nojiri} (please, see footnote~\ref{footnote1}). In the particular situation where $\alpha=1/(2p)-1/2$, being $p$  a positive integer, all the derivatives of $H$ are well defined at $t_{max}$ and consequently there is no singularity at $a_{\rm{max}}$.

\begin{figure}[h]
\begin{center}
\includegraphics[width=6cm]{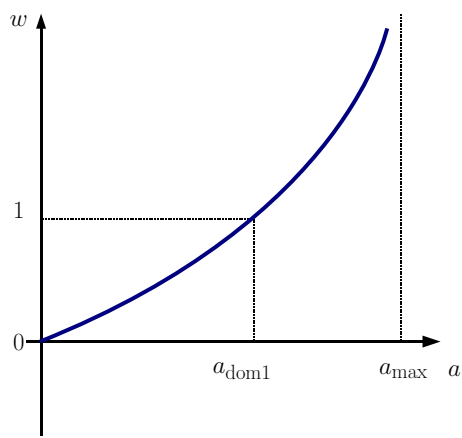}
\end{center}
\caption{Plot for the pressure$/$energy density ratio; $w$, for GCG~I and GCG~III, as a function of the scale factor. The dominant energy condition is not fulfilled for a scale factor larger than $a_{\rm{dom1}}$.}
\label{adom1f}
\end{figure}

Before concluding this subsection, we would like to point out that the universe never accelerates in this case. In fact, $w\equiv p/\rho$ is always positive. Therefore, the null, strong  and weak energy conditions are always fulfilled. Nevertheless, the dominant energy condition is  violated for scale factor larger than $a_{\rm{dom1}}$ where
\begin{equation}
a_{\rm{dom1}}=2^{-\frac{1}{3(1+\alpha)}}a_{\rm{max}};
\label{adom1}\end{equation}
consequently, the dominant energy condition is violated at the sudden singularity (see Fig.~\ref{adom1f}).

%%%%%%%%%%%%%%FIG4%%%%%%%%%%%%%%%
%%%%%%%%%%%%%%%%%%%%%%%%%%%%%%%%

\subsection{GCG II}\label{B}

\begin{figure}[t]
\begin{center}
\includegraphics[width=6cm]{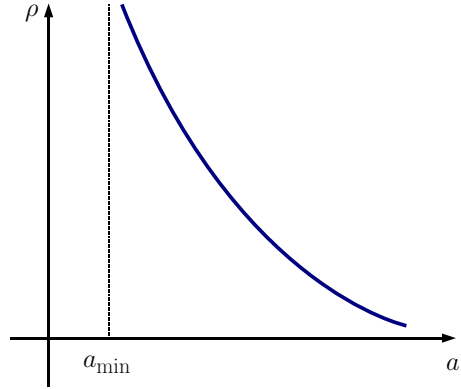}
\end{center}
\caption{A schematic plot of a past big freeze singularity.This singularity takes places at a finite past cosmic time where not only the energy density but also the pressure blows up. Therefore, both the Hubble rate and its cosmic time diverge as well at the big freeze event.}
\label{bbigfree}
\end{figure}

We briefly analyse what happens to a homogeneous and isotropic universe filled with a GCG II; i.e. with $A<0$, $B>0$ and $1+\alpha < 0$. It turns out that the scale factor is bounded from below, $a_{\rm{min}}\leq a$, where the  minimum scale factor is given by Eq.~(\ref{tres}). The GCG energy density as well as its pressure both diverge at $a_{{\rm min}}$, leading to a past curvature singularity (cf. Eq.~(\ref{R})). It corresponds to  a big freeze singularity or a Type III singularity  in the notation of Ref.~\cite{Nojiri} (see Fig.~\ref{bbigfree}). This singularity takes places at a finite past (see Eqs.~(\ref{cuatro}) and (\ref{cinco}) which also apply to in this situation). After the big freeze  singularity, the energy density  $\rho$ keeps decreasing until it vanishes at very large value of the scale factor where the GCG has a dust-like behaviour.

%%%%%%%%%%%%%%%FIG5%%%%%%%%%%%%%%%%%%%%
%%%%%%%%%%%%%%%%%%%%%%%%%%%%%%%%%%%%%%%%

\begin{figure}[h]
\begin{center}
\includegraphics[width=6cm]{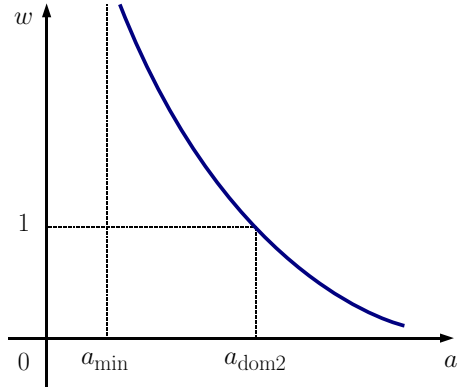}
\end{center}
\caption{Plot of the ratio of the pressure and the energy density; i.e. $w$, for GCG~II and GCG~IV, as a function of the scale factor. The dominant energy condition is not fulfilled for a scale factor smaller than $a_{\rm{dom2}}$.}
\label{adom2f}
\end{figure}

We would like to point out that, similar to the previous case, a universe filled with this fluid would never accelerate. In fact, $w\equiv p/\rho$ is always positive. Therefore, the null, strong  and weak energy conditions are always fulfilled. However, the dominant energy condition is violated for scale factor smaller than $a_{\rm{dom2}}$, defined as 
\begin{equation}
a_{\rm{dom2}}=2^{-\frac{1}{3(1+\alpha)}}a_{\rm{min}},
\label{adom2}\end{equation}
which implies that  the dominant energy condition is violated at the big freeze singularity (see Fig.~\ref{adom2f}).

%%%%%%%%%%%FIG6%%%%%%%%%%%%%%%%%%%%%
%%%%%%%%%%%%%%%%%%%%%%%%%%%%%%%%%%%

\subsection{GCG III and IV}\label{C}

Finally, we  analyse a universe filled with a GCG such that  $A>0$, $B<0$ and $1+\alpha=1/(2n)$ where $n$ is an integer. 

Let us consider that the fluid is characterised by a positive $n$ which we will refer to as GCG III. Then, the scale factor is bounded from above by $a_{{\rm max}}$ introduced in Eq.~(\ref{seis}).  At small scale factors the fluid behaves like ``dust''; i.e. $w\sim 0$ although $p$ diverges, and therefore a big bang singularity shows up.  Consequently, a FLRW universe filled with such a  fluid  would start at a curvature singularity. At  $a=a_{{\rm max}}$, the energy density and the pressure of GCG III vanish and $w$ acquires very large values, although the scalar curvature is well defined (see Eq.~(\ref{R})). Using Eqs.~(\ref{amaxappr1})-(\ref{amaxappr2}) which hold, it can be proven that it will take an infinite cosmic time for a universe filled with this fluid to reach its maximum size. As it can be realised, the advantage of having a  quantised  $\alpha$ is to remove the possible different future singularities which may be induced by GCG I. After all, this advantage is again not that surprising as having  $n>0$ implies $-1<\alpha\leq -1/2$, precisely a situation without future singularities for GCG I. Despite the absence of a future singularity at $a_{{\rm max}}$ and the fulfilment of the null, strong  and weak energy conditions, the dominant energy condition is violated for scale factors larger than $a_{\rm{dom1}}$ (see Eq.~(\ref{adom1}) and Fig.~\ref{adom1f}).

For completeness, we briefly analyse what happens if $n < 0$ a case which we name GCG IV. The scale factor is then bounded from below; i.e. $a_{\rm{min}}\leq a$, where $a_{{\rm min}}$ was defined by Eq.~(\ref{tres}). At the minimum scale factor, $\rho$ and $p$ (as well as $w$) diverge inducing consequently an initial  big freeze  singularity (see Fig.~\ref{bbigfree}) because the singularity happens at a  finite past cosmic time (Eqs.~(\ref{cuatro}) and (\ref{cinco}) both apply after replacing $A(1+\alpha)$ by $|A(1+\alpha)|$). Afterwards, the universe starts expanding although never accelerating, and therefore whereas the null, strong  and weak energy conditions are satisfied, the dominant energy condition is violated at radii smaller than $a_{\rm{dom2}}$ defined by Eq.~(\ref{adom2}) (see Fig.~\ref{adom2f}). On the other hand, at large scale factor GCG IV mimics a dust fluid.

\begin{table}[ph]
\tbl{This table summarises the asymptotic behaviour of a universe filled with the four types of GCG analysed in the text.
The future asymptotic behaviours labelled by (1) and (3) corresponds to  $-1<\alpha\leq-1/2$ and $0<\alpha$, respectively. The future asymptotic behaviour labelled by (2) corresponds to $-1/2<\alpha< 0$, where $\alpha$ cannot be expressed as $\alpha=1/(2p)-1/2$, with $p$ a positive integer. If  $-1/2<\alpha< 0$ and $\alpha$ can be expressed as $\alpha=1/(2p)-1/2$, with $p$ a positive integer, there is no future singularity and the universe reaches its maximum radius at a finite future cosmic time.}
{\begin{tabular}{|c|c|c|c|c|c|}
\cline{1-6}
A,B	&  $1+\alpha$	&$a$	&$\rho$	                          &Past	                               &Future\\
\cline{1-6}
	&	        &	&                   &  & (1) no singularity/infinite future     \\
\cline{6-6}
$A<0$ & positive	&$0\leq a<a_{{\rm max}}$	&  $0\leq\rho<\infty$                 &  dust-like       &(2) type IV singularity \\
\cline{6-6}
&	&	&	& 	& (3) sudden singularity \\
\cline{2-6}
$B>0$	& negative	&$a_{{\rm min}}\leq a <\infty$	&$0\leq\rho<\infty$ 	&big freeze singularity 	&dust-like\\
\cline{1-6}
$A>0$	&$(2n)^{-1}>0$	&$0\leq a<a_{{\rm max}}$	&$0\leq\rho<\infty$ &dust-like 		&no singularity/infinite future\\
\cline{2-6}
$B<0$	&$(2n)^{-1}<0$	&$a_{{\rm min}}\leq a <\infty$	&$0\leq\rho<\infty$ 	&big freeze singularity	&dust-like\\
\cline{1-6}
\end{tabular}\label{table3}}
\end{table}

\section{Discussion and conclusions}

In this work we have analysed in some detail a GCG-filled universe, pointing out the different kinds of singularities that may face in its future or past (see the tables for a summary of our results). In particular, we have shown that, although there are no big rip singularities in GCG models, other kinds of singularities can take place for other more complicated models. The emergence of some of this singularities, in particular the big freeze singularity, could be thought to be caused by the non-fulfilment of the null, strong and weak energy conditions \cite{Visser:1995cc} by a phantom GCG\footnote{Please, see footnote \ref{footnote4}.}. For this reason, we have studied models of GCG satisfying the above mentioned energy conditions. We have then shown that even in this case a sudden \cite{Barrow:2004xh}, a past big freeze and  a type IV singularity, in the notation of Ref.~\cite{Nojiri}, could surface. Therefore, the fulfilment of the null, strong and weak energy conditions does not avoid the appearance of such singularities, particularly the big freeze one. We have also shown that in this case the dominant energy condition is not fulfilled at these sort of singularities (see Figs.~\ref{adom1f} and \ref{adom2f}). 

In section \ref{DPGCG} we have shown how a brane can introduce a cut off for the energy density, for which case it modifies the model in such a way that the universe become always finite in the past (see Table II). As it is expected, brane effects can generally be relevant at sufficiently early times and, in fact, PGCG and DPGCG have late times behaviour which are rather indistinguishable, so that the universe essentially will become either asymptotically de Sitter or it will show a future big freeze singularity (see Table I and II).

Before closing up, we should remind that close to a big rip singularity or a big brake event, the classical space-time breaks down and therefore a quantum analysis is required \cite{Dabrowski:2006dd}. We similarly expect that quantum effects became also important as one approaches a big freeze singularity. In particular, the kind of past or future big freeze singularities we have pointed out  in this paper need to be analysed from a quantum-mechanical point of view \cite{Bouhmadi-Lopez:2004mp}. As a starting point, it  would be interesting to perform such an analysis in the cases of the GCG II or GCG IV models, introduced in section \ref{GCG}, where at least we know that the null, strong and weak energy conditions are satisfied. We hope to report on this interesting issue in the near future.

\section*{Acknowledgements}

MBL is  supported by the Portuguese Agency Funda\c{c}\~{a}o para a Ci\^{e}ncia e
Tecnologia through the fellowship SFRH/BPD/26542/2006 and research grant FEDER-POCI/P/FIS/57547/2004.
PMM gratefully acknowledges the financial support provided by the I3P framework of CSIC and the European Social Fund. This research was supported in part by MEC under Research Project No. FIS2005-01181.

\end{document}